\documentstyle[prl,aps,epsf,times,float,graphicx]{revtex}
\graphicspath{{/home/janko/Papers/Neutron/}}
\DeclareGraphicsExtensions{.eps,.ps}
\begin{document}
\draft
\twocolumn[\hsize\textwidth\columnwidth\hsize\csname@twocolumnfalse\endcsname
\title{Thermodynamic Constraints on the Magnetic Field Dependence of
  the Neutron Resonance in Cuprate Superconductors}
\author{Boldizs\'ar Jank\'o}
\address{Materials Science Division, Argonne National Laboratory \\
9700 South Cass Avenue, Argonne, Illinois 60439, USA} 
\date{\today} \maketitle
\begin{abstract}
  In order to test the recently proposed connection between the
  neutron resonance and the specific heat anomaly in cuprates, the
  experimental specific heat data on $YBa_2Cu_3O_{6.93}$ and a
  theoretical estimate of the single particle fermionic contribution
  to the specific heat is used to provide an upper bound for the
  intensity of the neutron peak. The deduced peak intensity is similar
  in magnitude and temperature dependence to that observed in neutron
  scattering experiments, and it is constrained to decrease strongly
  under the influence of moderate magnetic fields oriented along the
  c-axis of the crystal. An explanation is proposed for the predicted
  suppression, based on the observation that the resonance intensity
  is very sensitive to superconducting phase correlations. 
\end{abstract}
]
\makeatletter
\global\@specialpagefalse
\def\@oddhead{REV\TeX{} 3.0\hfill MSD-ANL Theory Group Preprint}
\let\@evenhead\@oddhead
\makeatother

The resonance observed in inelastic neutron scattering remains one of
the most heavily debated features of cuprate superconductivity.  Its
initial observation in $YBa_2Cu_3O_{6+x}$ \cite{Rossat} sparked
numerous experimental\cite{exp} and theoretical
\cite{Lavagna,Quinlan,Demler,Levin,Mazin,Anderson,Abrikosov,Pines,Lee}
investigations. The detection of the resonance in
$Bi_2Sr_2CaCu_2O_{8+\delta}$ \cite{Keimer}, together with several
recent studies suggesting a direct link between the resonance and
salient features of other experiments \cite{Demler,Norman,Carbotte},
give further support to the suspicion, that the neutron resonance is a
key phenomenon in high $T_c$ superconductors.

Among the possible links, the one between the thermodynamic and
magnetic anomalies deserves special attention. Taking a general
argument by Scalapino and White\cite{Doug} further, Demler and
Zhang\cite{Zhang} pointed out that the emergence of the neutron
resonance below the superconducting transition $T_c$ results in a
decrease in exchange energy that could account for the entire
superconducting condensation energy. They proposed a simple relation
between the resonance intensity $I_{\rm res} (T)$ and the
corresponding resonance contribution $C_{\rm res} (T) $ to the
specific heat:
\begin{equation}
C_{\rm res}(T) = - \frac{J}{2\mu_B^2} \Bigl(\frac{d I_{\rm res} (T)}{dT}\Bigr).
\label{eq:dz}
\end{equation}
Here $\mu_B$ is the Bohr magneton and $J$ is the nearest-neighbor
spin-spin coupling. Dai and his collaborators \cite{Dai} calculated
$\partial_T I_{\rm res} (T)$ from neutron scattering data on
$YBa_2Cu_3O_{6+x}$, used Eq.(\ref{eq:dz}) to compute the specific heat
anomaly corresponding to the neutron resonance, and
compared the outcome to the data of Loram and his
collaborators\cite{Loram}. They concluded with the remarkable
observation that the data are compatible with Eq.(\ref{eq:dz}).

\begin{figure}
  \centerline{\includegraphics[width=2.5in]{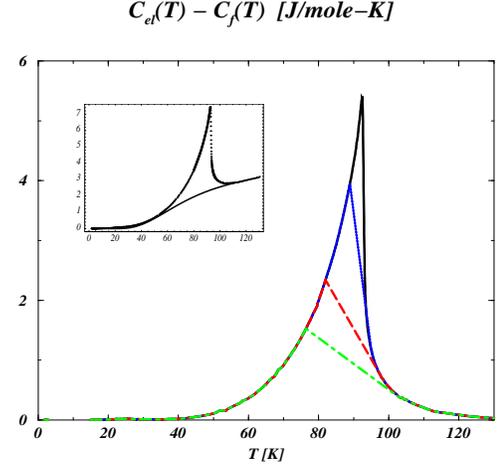}} 
\vspace*{3ex}
\caption{{\bf Inset:}The measured specific heat $C_{\rm el} (T)$ (dots), and
  the lower bound for the fermionic contribution $C_f(T)$ (solid
  line), obtained by numerical evaluation of Eq.({\protect\ref{cf}}),
  using an s-wave gap $\Delta = 25\,{\rm meV}$ at all temperatures.  The
  normal state density of states was fitted to reproduce the normal
  state data near $T_0$.  {\bf Main figure}: The anomalous part,
  obtained by subtracting the calculated lower bound, $C_f(T)$, from
  the experimental data, $C_{\rm el}(T)$ .  The effect of the magnetic
  field along the {\em c-axis} of the crystal is simulated by using
  the results of Ref.  \protect\cite{Junod}. The peak of the anomaly
  is gradually shifted to lower temperatures and reduced: $100 \% $ and
  93 K (zero field, solid curve), to $71\% $ and 89 K ($B = 1\, {\rm
    T}$, dotted curve), $45 \%$ and 81 K ($B = 3\, {\rm T}$, dashed
  curve), and finally $30 \%$ and 75 K ($B = 6\, {\rm T}$, dash-dotted
  curve)}
\label{fig:cvh}
\end{figure}

However, if Eq.(\ref{eq:dz}) holds, then its inverse must also be valid.
\begin{equation}
  I_{\rm res} (T) = (2\mu_B^2/J) \int^{T_0}_T dt C_{\rm res}(t),
\label{eq:inverse}
\end{equation}
where $T_0$ is an upper cutoff temperature, chosen to be larger than
$T^*$, the resonance onset temperature. Thus, it should be possible to
check the link between the thermodynamic and magnetic measurements by
starting from the specific heat measurements instead.  After only a
qualitative examination of typical specific heat data $C_{\rm el} (T)$
(see, for example inset of Fig. \ref{fig:cvh}) one concludes that
performing such a task is not possible without an adequate account of
the fermionic contribution to the specific heat.  In fact, the
specific heat jump in conventional superconductors is {\em entirely}
due to the onset of a finite gap in the quasiparticle spectrum
\cite{Rickayzen}.

More important is the observation that if $C_{\rm res}$ is defined
adequately, the inverted relation Eq.(\ref{eq:inverse}) allows one to
impose a severe constraint on $I_{\rm res}(T)$. It is known for some
time, that the specific heat anomaly is removed
\cite{Ginsberg,Junod,Lawrie,Carrington,Junod2} by a magnetic field of
several Tesla oriented along the c-axis of the cuprate crystal. This
means that Eq. (\ref{eq:inverse}), used with thermodynamic data taken
in magnetic field, predicts large field induced effects on the
resonance intensity.  This is a new property of the neutron resonance,
that has not been observed yet. The aim of this paper is therefore to
point out that $(a)$ an {\it upper bound} can be given for $C_{\rm
  res} (T)$ in $YBa_2Cu_3O_{6.93}$, which in turn yields a reasonable
account of $I_{\rm res} (T)$, $(b)$ the magnetic field sensitivity of
$C_{\rm res} (T)$ {\it constrains} $I_{\rm res} (T)$ to be strongly
field dependent, and $(c)$ an intimate connection exists between the
presence of a sharp resonance and superconducting long range order.
This last observation leads to a natural explanation of the predicted
field-induced suppression in resonance intensity.

The procedure for obtaining an upper bound for $C_{\rm res}(T)$ is
based on the observation that one can give a lower
bound for the specific heat contribution $C_{\rm f}(T)$ of single
particle fermionic excitations.  The fermionic specific heat
contribution can be calculated from $\Omega_f (T,V)$, corresponding to
the fermionic part of thermodynamic potential\cite{agd} 
\begin{equation}
  \Omega_f (T,V) = - \frac{1}{\beta} {\rm Tr} \sum_{{\bf k},\ell,\sigma}
  \Big\lbrace \ln [(G_{{\bf k},\ell})^{-1}] - \Sigma_{{\bf
      k},\ell} G_{{\bf k},\ell}
  \Big\rbrace,
\end{equation}
where the single particle fermion propagator has the Nambu form
$G_{{\bf k},\ell} = (i\zeta_\ell - \epsilon_{\bf k} \tau_3 -
\Delta_{\bf k} \tau_1)^{-1}$, corresponding to the
Bardeen-Cooper-Schrieffer (BCS) self energy $\Sigma_{{\bf k},\ell} =
\Delta_{\bf k}\tau_1$;$\tau_i $ are Pauli matrices, $ \zeta_\ell =
(2\ell +1 ) \pi k_BT$, and
$\Delta_{\bf k} (T) $ is the superconducting gap. After performing the
trace over the Pauli matrices, and the Matsubara sum, one obtains
\begin{equation}
  \Omega_f (T,V) =  \sum_{{\bf k},\sigma}
  \Big\lbrace \frac{2}{\beta} \ln [1 - f_{\bf k}] - E_{\bf k}
  + \frac{\Delta^2_{\bf k}}{2 E_{\bf k}} (1 - 2f_{\bf k})
  \Big\rbrace,
\label{Omega}
\end{equation}
where $f_{\bf k} = [1 + \exp (\beta E_{\bf k})]^{-1}$ is the Fermi
function.  This is, not surprisingly, the usual form of the BCS
thermodynamic potential \cite{Rickayzen}, since in a conventional
superconductors the elementary excitation are fermionic in nature.

The same statement does not necessarily hold for the cuprate
superconductors. It was argued\cite{Zhang} that the specific heat
jump in the cuprates is in large part due to the appearance of new
bosonic degrees of freedom below $T_c$, The degradation\cite{Loram} of
the specific heat jump into an anomalous cusp in the underdoped regime
adds further support to the possibility that jump is caused by the
presence of bosonic excitation in the superconducting state.  The
expression in Eq. (\ref{Omega}) will be used to separate from $C_{\rm
  el}$ the part that could, in principle, be due to bosonic
excitations. This will be done by subtracting from $C_{\rm el}$ the
fermionic contribution, computed from Eq. (\ref{Omega}) and the
thermodynamic relation $C_{\rm f} (T) = - T \partial^2 \Omega_f
(T,V)/(\partial T)^2$. There is no information available on the
temperature dependence of $\Delta_{\bf k}(T)$ in $YBa_2Cu_3O_{6+x}$.
Under these circumstances it is more useful to provide a controlled
lower bound for the fermionic specific heat. This is possible to
achieve by taking the low temperature gap value $\Delta \simeq 25\,
{\rm meV}$ from tunneling measurements\cite{YBCOgap}, and calculating
$C_{\rm f}(T)$ under the assumption that $\Delta$ is {\em constant at
  all temperatures}. In this way one obtains a lower bound for the
fermionic specific heat contribution, as compared to the case when
$\Delta (T)$ is decreasing with increasing temperature. Clearly the
constant gap model systematically {\em underestimates} $C_{f}(T)$ in a
system such as $YBa_2Cu_3O_{6.93}$, where $\Delta (T)$ is known to
close near $T_c$.  In the underdoped $YBa_2Cu_3O_{6+x}$ samples, due
to the onset of the pseudogap, the constant gap model becomes an even
better approximation\cite{Ding}.  Unfortunately, no tunneling data is available
yet for such crystals.

Within the constant gap model Eq. (\ref{Omega}) gives the following
result for the fermionic specific heat
\begin{eqnarray}
C_f(T) = \frac{k_B}{2} \sum_{\bf k} 
{\rm sech}^3\bigl(\frac{\beta E_{\bf k}}{2}\bigr)
\Bigr[ \bigl( \beta \epsilon_{\rm k} \bigr)^2 {\rm cosh}
\bigl(\frac{\beta E_{\bf k}}{2}\bigr) \nonumber \\
+ \bigl( \beta^3 \Delta^2_{\bf k} E_{\bf k}\bigr)
\sinh \bigl(\frac{\beta E_{\bf k}}{2}\bigr) \Bigl]
\label{cf}
\end{eqnarray}

The anomalous part, $C_{\rm res} (T)$, will be defined as the
difference in the measured \cite{Loram} electronic specific heat
$C_{\rm el} (T)$ and the fermionic contribution
\begin{equation}
  C_{\rm res}(T) \simeq C_{\rm el}(T) - C_{\rm f}(T).
\label{cres}
\end{equation}
The specific heat data of Loram {\it et al.} for $YBa_2Cu_3O_{6.93}$,
displayed in in the inset of Fig.\ref{fig:cvh}, together with Eqs.
(\ref{eq:inverse}),(\ref{cf}) -(\ref{cres}), can now be used to compute
the temperature dependence of the neutron resonance, as predicted by
the thermodynamic data.  
At low temperatures this procedure gives
$I_{\rm res} (0) \sim 0.01\, \mu^2_B{\rm f.u.}^{-1}$, to be compared
to the corresponding experimental value of $0.06\, \pm 0.04
\mu^2_B{\rm f.u.}^{-1}$, where ${\rm f.u.}$ stands for formula unit.

\begin{figure}
  \centerline{\includegraphics[width=3in]{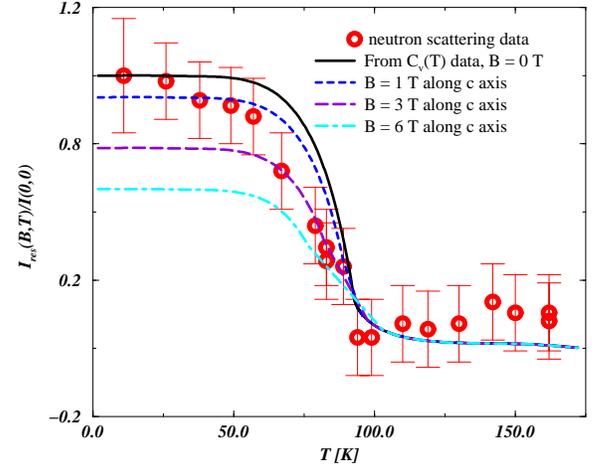}} 
\vspace*{3ex}
\caption{The neutron resonance intensity for the optimally doped 
  compound $YBa_2Cu_3O_{6.93}$: Dots are the zero field results of
  Ref. {\protect\cite{Dai}}; lines were derived from specific heat
  data of Refs.{\protect\cite{Loram,Junod}} as explained in the text.
  All the curves were normalized with the zero field, low temperature
  value of the deduced intensity, $0.01 \mu^2_B{\rm f.u.}^{-1}$.}
\label{fig:normintens}
\end{figure}

While the deduced intensity is still the same order of magnitude as
the experimental value of Ref.  \cite{Dai}, the agreement is less
striking. On the other hand, the fact that the resonance intensity
required by the thermodynamic changes is smaller than that observed in
neutron scattering supports the possibility that the source of
superconducting condensation energy is in fact the gain in exchange
energy, as suggested by Demler and Zhang\cite{Zhang}. The deduced
$I_{\rm res} (T)$ is somewhat affected by ambiguities in choosing
$T_0$, by experimental errors in the tunneling gap $\Delta$, and in
the specific heat data itself, introduced by the necessary phonon
subtraction.  The normalized zero field intensity is shown as the
solid curve in Fig.  \ref{fig:normintens}, together with the
experimental data from Ref.~\cite{Dai}. The agreement is satisfactory
except near the transition temperature $T_c \sim 92.5 \, {\rm K}$,
where the intensity is overestimated. This can be considered an
indirect signature of temperature dependence of the superconducting
gap.

As pointed out by several groups \cite{Ginsberg,Junod,Lawrie}, the
specific heat anomaly near $T_c$ is removed in $YBa_2Cu_3O_{6.93}$ by
a magnetic field ${\rm B} \sim 10\, T$ along the c axis, while $C_{\rm
  el}(T)$ away from $T_c$ is unaffected.  This is highly unusual,
since the applied magnetic field is only a small fraction of the upper
critical field  $B_{\rm c2} \sim 100\,{\rm T}$.  Similar
effects have been recently documented in other cuprate materials as
well \cite{Carrington,Junod2}. The recent work of Junod and coworkers
\cite{Junod2} presents an especially detailed account of the field
dependence of the specific heat anomaly. They contrast the behavior
seen in the cuprates with that in a low $T_c$ superconductor (Nb-Zr
alloy), where the anomaly is pushed to lower temperatures, but never
removed. Indeed, the behavior seen in cuprates seems to go beyond the
conventional framework, according to which as long as $T_c$ is finite,
the specific heat jump is also finite\cite{Fetter}.

The most detailed set of data for the field dependence of the anomaly
in $YBa_2Cu_3O_{6.93}$ was presented in Ref. \cite{Junod}.  In these
measurements, however, the phonon background was not subtracted, as in
the data Loram {\it et al.} used above. In order to avoid subtraction problems,
the finite field specific heat jumps were obtained by imposing on
Loram and coworkers' zero field data \cite{Loram} the suppression in
the height and the shift towards lower temperatures of the specific
heat maximum observed by Junod's group \cite{Junod}. The result is
shown in the main part of Fig. \ref{fig:cvh}. 

When field-suppressed specific heat anomalies are used in
Eq.(\ref{eq:inverse}), the resulting temperature dependence of the
deduced resonance intensity deviates from what was measured in zero
field, as shown in Fig. \ref{fig:normintens}. In fact, these results
show that $I_{\rm res} (B,T)$ should be depleted at low temperatures
beyond the present experimental error when $ B > 6\, T$ (see figure).
Bourges {\it et al.} \cite{Bourges} measured the neutron resonance in
magnetic field and found $I_{\rm res} (T\sim 6\,{\rm K})$ to be {\it
  insensitive} to fields of order $ B \sim\,12\,T$ oriented along the
$ab$ plane of the sample.  Their result does not invalidate the claim
of Dai {\it et al.}, since in that geometry the specific heat anomaly
is only moderately depleted, and not removed by the field\cite{Junod}.
However, the measurement of $I_{\rm res} (B,T)$ with ${\bf B}$ along
the c-axis seems to be a crucial experiment for not only establishing
the validity of Eqs.(\ref{eq:dz}) and (\ref{eq:inverse}), but also to
gain important insight into the nature of the neutron resonance itself.

The strong magnetic field dependence of the resonance predicted by the
above procedure raises immediately a key question: What is the main
mechanism responsible for inducing such a large change in the neutron
intensity? In the remainder of this paper I will argue that there is
an intimate connection between the resonance and correlations in the
phase of the superconducting order parameter. This connection, besides
offering a natural explanation for the field sensitivity of the
resonance, also seems to have direct relevance to some puzzles in the
zero field properties of the resonance, like the existence of the
resonance in bilayer materials only\cite{exp}, and the presence of a
broad precursor resonance in the pseudogap state\cite{Dai}.

There are three important clues to consider. First, irrespective of
doping, a sharp resonance is observed only below $T_c$, after off
diagonal long range order is established. Second, the magnetic field
required to affect the resonance is predicted by the above analysis to
be much smaller than $B_{\rm c2}$. Scanning tunneling spectra
\cite{STM} show that such fields do not broaden out the
superconducting singularities significantly. Finally, a c-axis field
strongly affects the specific heat jump and the resonance, but only
mild changes are observed if the field is along the ab-plane
\cite{Junod}, consistent with the results of Bourges {\it et
  al.}\cite{Bourges}. The strong dependence on the field direction
suggests that Zeeman splitting is irrelevant and that the resonance is
not affected by shielding currents along the c-axis itself, when the
field is parallel to the $CuO_2$ planes.  On the contrary, the
resonance is affected considerably by the presence of {\it in-plane}
shielding currents circulating around the cores of Abrikosov vortices,
when the field is parallel to the c-axis. As discussed in more detail
below, this translates to the statement that in-plane long range
coherence as well as short range c-axis coherence (intra-bilayer) does
matter for the resonance, whereas long range c-axis phase coherence
does not.

In order to make these observations more transparent, it is instructive
to examine the bare spin susceptibility, where the phase dependence is
indicated explicitly \cite{Ioffe}
\begin{eqnarray}
\chi^0({\bf Q},i\Omega_n) = \chi^0_{GG} ({\bf Q}, i\Omega_n)
\ \ \ \ \ \ \ \ \ \ \ \ \ \ \ \ \ \ \ \ \ \ \ \ \ \ \nonumber \\
+ \frac{1}{\beta}\sum_{{\bf p,}m} \chi^0_{FF}({\bf
    Q-p},i\Omega_n-i\omega_m) \Lambda ({\bf p}, i\omega_m).
\label{Lambda}
\end{eqnarray}
Here $\chi^0_{GG}({\bf Q}, i\Omega_n)$ and $\chi^0_{FF}({\bf Q},
i\Omega_n)$ are the normal and superconducting contributions to the
bare susceptibility, $\chi_{FF}^0({\bf q},\omega_n ) =
\frac{1}{\beta}\sum_{{\bf k},\ell} F_{{\bf
    k+q},}(\zeta_\ell+\omega_n)F_{{\bf k}}(\zeta_\ell)$, and $\Omega_m
= 2m\pi k_BT$. A similar expression holds for $\chi^0_{GG}({\bf Q},
i\Omega_n)$, but with the Gor'kov propagators replaced by the diagonal
part of the superconducting Green's function. The function $\Lambda
({\bf p},i\omega_m)$ is the Fourier and Matsubara transform of the charge
displacement correlation function\cite{Phase}, $\Lambda ( {\bf r-r'},
\tau - \tau') = \langle T_\tau [e^{i\phi(r,\tau)}
e^{-i\phi(r',\tau')}] \rangle$, and $\phi(r,\tau)$ is the phase of the
superconducting order parameter at position $r$ and imaginary time $\tau$.

Equation (\ref{Lambda}) can provide new insights even for the zero
field case. In the absence of phase fluctuations, corresponding to the
BCS mean field level, $\Lambda ({\bf q}, i\omega_m) = \beta \delta
({\bf q}) \delta_{m,0}$, and one obtains the standard result used as a
starting point in essentially all theoretical descriptions of the
resonance
\cite{Lavagna,Quinlan,Demler,Levin,Mazin,Anderson,Abrikosov,Pines,Lee}.
In zero external field the phase fluctuations are uniform in space and
Gaussian around the mean phase $\phi_0$, and consequently the
correlator $\Lambda ({\bf p},\omega)$ is still dominated by its low
frequency, long wavelength value $\Lambda (0,0) = \exp[-\langle \delta
\phi^2 \rangle]$. Thus, phase fluctuations reduce the contribution of
the anomalous susceptibility $\chi_{FF}^0$, and consequently the
resonance
intensity\cite{Lavagna,Quinlan,Demler,Levin,Mazin,Anderson,Abrikosov,Pines,Lee},
since neither the bare, nor the RPA-dressed $\chi_{GG}$ can account on
its own for the resonance. In single layer materials, due to the weak
interplane coupling and strong two dimensionality, quantum phase
fluctuations can be quite large, and consequently $\Lambda$ very
small. In contrast to single layer materials, the strong Josephson
coupling between bilayers can stabilize the superconducting state with
a rigid phase. This could be the main reason why the neutron resonance
has only been observed in bilayer materials, and not in single layer
materials, such as $La_{2-x}Sr_xCuO_4$\cite{exp}. The broad maximum
observed\cite{Dai} in the {\em pseudogap regime} of underdoped
$YBa_2Cu_3O_{6+x}$ could be the precursor of the resonance below
$T_c$, provided that strong pairing fluctuations are present in this
regime \cite{Maly}.

In the Abrikosov vortex phase induced by a field along the c-axis, the
weight of $\Lambda ({\bf p}, \omega)$ is no longer concentrated at
long wave lengths and low frequencies. As shown by Glazman and
Koshelev \cite{GK}, the vortex lattice destroys the off-diagonal long
range order in the planes at finite $T$, while long range order along
the field direction is broken only at elevated temperatures. For the
in plane correlations, they find $\lim_{r \rightarrow \infty} \Lambda
(r, z=0) = \lim_{r \rightarrow \infty} \exp[-r/\ell(B)] = 0 $, where
$\ell(B) \propto B^{-1/2}$ is the in-plane phase correlation length in
magnetic field.  Thus, the field will induce a momentum broadening
$\delta {\bf p} \propto \sqrt{B}$ in the resonance, and a reduced
intensity.  In general, a form of $\Lambda ({\bf q},\omega)$ that
deviates from the singular mean field result, will upset the resonance
condition satisfied by the zero field susceptibility for
$Q=(\pi,\pi,\pi); \omega = \omega_0$.

The arguments presented above are quite general and consequently have
direct relevance not only to the magnetic field dependence of the
neutron resonance, but also to the finite field behavior of optical
conductivity, tunneling density of states, Raman scattering cross
section, and other spectroscopic and thermodynamic properties of
cuprate superconductors. A detailed study of all these effects is
currently under way, and will be reported in a later publication.

I am grateful to Dr. John W. Loram for making his specific heat data
available. I would also like to thank Prof. A. A. Abrikosov, Dr.  Joel
Mesot, and especially Dr. Michael R. Norman for numerous discussions,
and to Dr. Philippe Bourges, Dr. Pengcheng Dai and Dr. Herbert A. Mook
for correspondence.
This research was supported in part by the NSF under awards
DMR91-20000 (administrated through the Science and Technology Center
for Superconductivity), and the U.S.  DOE, BES, under Contract
No.~W-31-109-ENG-38.


\end{document}